# Detecting Sybil Attacks in Vehicular Ad Hoc Networks

Salam Hamdan

Computer Science Department

Princess Sumaya University for technology

Amman, Jordan

S.hamdan@psut.edu.jo

Amjad Hudaib

Computer Science Department

University of Jodan

Amman, Jordan

AHUDAIB@JU.EDU.JO

Arafat Awajan

Computer Science Department

Princess Sumaya University for technology

Amman, Jordan

[AWAJAN@PSUT.EDU.JO](mailto:AWAJAN@PSUT.EDU.JO)

**Abstract**

Ad hoc networks is vulnerable to numerous number of attacks due to its infrastructure-less nature, one of these attacks is the Sybil attack. Sybil attack is a severe attack on vehicular ad hoc networks (VANET) in which the intruder maliciously claims or steals multiple identities and use these identities to disturb the functionality of the VANET network by disseminating false identities. Many solutions have been proposed in order to defense the VANET network against the Sybil attack. In this research a hybrid algorithm is proposed, by combining footprint and privacy-preserving detection of abuses of pseudonyms (P2DAP) methods. The hybrid detection algorithm is implemented using the ns2 simulator. The proposed algorithm is working as follows, P2DAP acting better than footprint when the number of vehicles increases. On the other hand, the footprint algorithm acting better when the speed of vehicles increases. The hybrid algorithm depends on encryption, authentication and on the trajectory of the vehicle. The scenarios will be generated using SUMO and MOVE tools.

**Introduction**

VANET is a subgroup of Mobile ad-hoc network (MANET) in which the nodes are vehicles [1, 2]. VANETs are able to provide the communication between vehicles and infrastructure by a Dedicated Short Range Communication (DSRC) [3, 4]. However, VANET is vulnerable to different types of attacks [5, 6], Such as the Sybil attack, in which the intruder fabricates the identities of multiple vehicles [7]. The communication between entities in VANET is done through radio waves, and these waves contain a lot of private and sensitive information for applications, personal information of drivers and travelers. This information could be used to enhance roads safety and provide a comfortable driving for drivers. In the other hand, this information must be secured against intruders.

VANET network has three main system components which are the application unit (AU), on-board unite (OBU) and Road Side Units (RSU) [8]:

- Smart vehicle: Vehicles in VANET networks are equipped with OBU which is responsible on networking, processing, and determine vehicles locations and rout directions. Application unit (AU): This component is responsible many applications such as driver and vehicle safety by working as a warning device for safety application, also it works as a navigator for communication applications [9]
- On-Board unit: This device is used to make the inter-vehicle communication (V2V) and infrastructure to vehicles communication (V2R) [8]. This device is also used to communicate several types of message including safety oriented messages [9]
- Roadside unit (RSU): this device is physically located on fixed locations, such as traffic lights [9]. This device is used to expand communication range for the VANET network by exchanging messages with OBU's and broadcast important information to entities in its range, communicate with other RSU's, running safety application and provide OBU's with internet connection [8].

The infrastructure-less nature of VANET [10] and using broadcasting to transmit message, makes it vulnerable to different types of attacks [11, 12] such as Impersonation attacks [13], False attribute possession [14], Replay attack [15, 16], Tunneling attack [16], Message tampering [15], ID disclosure [17, 18] and the Sybil attack [19] in which the intruder fakes multiple identities [20]. This paper focuses on the Sybil attacks in VANETS.

Sybil node forges several identities such as pretending to be a police car, stealing vehicles identities or creating new identities. Sybil attack is a dangerous attack on VANET network because the attacker could initiate different types of attacks such as DoS attack to ruin communication between entities in the VANET network [1, 21]. Also, Sybil node could fabricate traffic jam in order to force other vehicles to take other routes. In addition, Sybil node could inject false information in the network and may put the life of passengers in danger [22].

Number of schemes have been proposed to detect and prevent Sybil attacks [5]. These schemes can be categorized into three classes: (1) resource testing, (2) position verification, and (3) encryption and authentication. Resource testing is based on counting the number of resources the node has such as its computing ability, communication bandwidth and so on [23, 24]. This scheme is considered inadequate to VANET since the smart vehicle can fabricate having more resources. Position verification, depends on linking one position with one identity. This scheme has many solutions. These solutions are discussed later. This research focuses on the encryption and authentication category that based on authentication mechanism and public key cryptography.

This research performs a Hybrid algorithm for the footprint and P2DAP schemes, by simulating this new hybrid scheme using ns2 simulator and generate the scenarios using special tools SUMO and MOVE tools. The rest of this paper is organized as follows; section II presents related work regarding Sybil attack detection schemes. Section III, discuss the proposed hybrid algorithms. Section IV, will evaluate the performance of the hybrid detection schemes. And finally, section V will conclude the paper.

**Related work**

In ad hoc network, Sybil attack detection schemes are classified into three classes, radio resource testing, identity registration and position verification based, and finally encryption and cryptography based schemes [23].

*Radio resource testing*

The first category is resource testing which is depends on the number of resources the node has, but in vehicular ad hoc network this category is inadequate, therefore, the malicious node can fake having more resources than legitimate node can have [25].

*Identity registration and position verification*

The second category is the position based category, this scheme is depends on linking one identity with one position, if the identity have more than one position then it might be a Sybil node [26]. Number of schemes has been proposed according to this scheme [27].

In [25] the authors proposed a scheme that is called Active Security through Seeing (PASS). There scheme depends on radars. In this scheme the radars are acting as the "eye" of the system, if the radar can see the node then it is exist, otherwise it is not. Although the transmission range of the radar is low, the nodes can exchange the information with the neighboring vehicles. In this scheme the authors assume that 85% of nodes are legitimate nodes which it is considered as a drawback. Also this approach suffers from long communication delays and long response time in addition to other potential security issues.

In [28] the authors proposed a scheme called tunable radar, which is an improvement on the previous scheme. By using the tunable radar instead of the static radar range, the range of radar can be changed by changing the sampling rate of the radar. The advantages of this scheme are to enhance the efficiency of detecting positions and also prevent potential position attacks in a local cell compared to PASS.

In [27] the authors proposed a solution that is called signal strength which depends on the signals that are captured by other vehicles in the network. In this scheme the nodes are classified into three classes' verifier, claimer and witness. A claimer periodically sends a beacon message that contains a claimed position. A witness node gathers the claimers beacon messages and estimates the position for each claimed node depending on the signal strength. A verifier node collects all received signal strength from the claimer nodes if the difference between the estimated position and the claimed position is large then it is a suspicious node.

*Defenses based on encryption and authentication*

These schemes depend on encrypting and decrypting the messages between vehicles, using symmetric, asymmetric, hash function or digital certificates [29] [30]. These schemes also ensure to link one position with one identity [5]. The following schemes are following under this category:

In [31] the authors proposed a scheme called footprint this scheme depends on the trajectory (path) of the vehicle to identify the vehicles. When a vehicle passes by an RSU, the RSU issues an authorization message to the vehicle, which is an evidence to prove that this vehicle passed by this RSU.

In [32] the authors proposed two schemes in their research depending on the type of the certificate. In their schemes they aim to reduce the system architecture requirements and the cost of the computational certificate managements. They also aim to make their schemes able to be deployed in an-early stage VANET. The components of their scheme are certificate authority (CA) and RSUs. The RSU is accountable on releasing the RSUs public keys certificates. This solution is depends on "the timestamp series certificate approach and the temporary certificate approach" [23], which are described as follows:

Series of timestamp certificate approach: A certificate that contains a current timestamp is generated by each RSU in the network. Each vehicle passes by an RSU will gain a certificate that contains the current timestamp, and according to this the vehicles will gain a series of timestamp certificates. The Sybil node can be detected if two vehicles have the same timestamps, because it is unlikely that two vehicles pass by a sequence of RSUs with the same timestamps. This approach will work in an inefficient way when applying in an urban environment and this is because of two challenges. The first challenge is the complexity of the roads. The second challenge is that in urban environment the road has many intersections and vehicles tend to slow down or stop at these intersections, so vehicles are likely to have similar timestamps, which will make Sybil node detection difficult. The solution to this issue is to deploy RSUs on the edges.

Temporary certificate approach: Each RSU generates key pairs on a temporary basis to be valid for a short period of time. A vehicle should be authenticated by an RSU in order to get the first certificate, after that, the vehicle regenerate its key pair and certificate with the following RSU resulting in a chained certificate. With this approach, the chances of detecting a Sybil node are higher when compared to the other approach that depends on one certificate at a time.

In [33] the authors proposed a scheme called "privacy-preserving detection of abuses of pseudonyms" (P2DAP). The system scheme components are the Department of Motor Vehicle (DMV) and Road Side Boxes (RSBs) which are the same as RSUs in previous schemes. RSBs are securely connected to the DMV via a backhaul wired network. Their scheme based on the assumption that DMV will yearly generates an adequate number of pseudonyms for all vehicles; these generated pseudonyms will be hashed in two steps, the first step using a one-way global key Kc; and this key will be distributed to all RSBs in the network, then from the hashed pseudonyms they selected a set of bits these selected bits are called "coarse-grained hash value", the pseudonyms are then organized according to these bits into groups, these groups are called "coarse-grained group". Afterwards these groups will be hashed again using another one-way key Kf known only by the DMV, and then group these pseudonyms according to selected bits

"fine grained hash value" into "fine-grained group". Each fine-grained group will be allocated to one vehicle. The Sybil attack detection will be done in two levels. First, RSBs overhear the message exchange and puts the used pseudonyms in a list; then the RSBs start to calculate the "course-grained hash value" for each event, if there are two vehicles or more are in the same "course-grained group" then it might be a Sybil attack. Therefore, the RSB will send a report to the DMV. Afterward the DMV will hash these suspicious pseudonyms using the Kc and then using the Kf, if these pseudonyms are in the same "fine-grained group" then they are Sybil nodes, otherwise it is a false alarm.

**Proposed algorithm**

In order to increase the performance and detection rate a hybrid algorithm is proposed to detect Sybil attacks. As mentioned before, P2DAP works better when the vehicle speed increases, while footprint works better with a larger number of nodes. Taking advantage of these results anew hybrid algorithm is proposed.

*Proposed algorithm components*

In VANET network, the communication between moving vehicles is happened via inter-vehicle communication and between vehicles and RSUs via Roadside-to-vehicle communications. Figure 1 shows the system components, which are (1) the smart vehicle, which is contained of on-board unit (OBU). (2) Road side units (RSU) which is responsible in providing the vehicles that passes by them with link-tags, provide the vehicles with their pseudonyms, ethically eavesdropping on the communication overhead and send a report to the DMV in the case of a suspicious event occurs. (3) Trust authority (TA) which is responsible for instituting trust among entities. (4) Department of motors and vehicles (DMV), which generates pseudonyms for vehicles yearly, grouping the pseudonyms and check whether the suspicious event whether it is a Sybil attack or a false alarm.

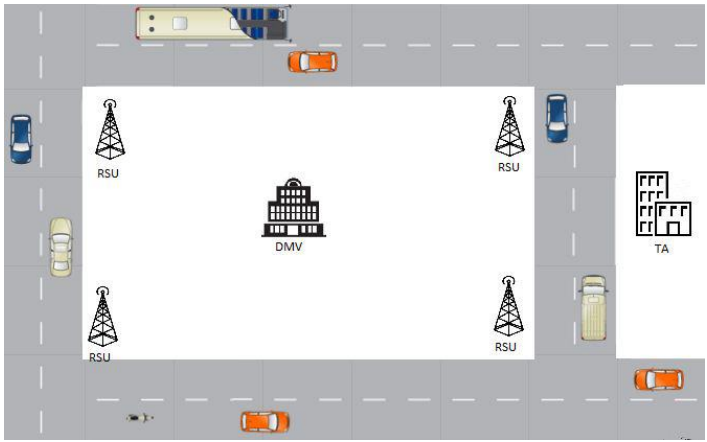

Figure 1: Proposed algorithm components

The proposed algorithm is a combination between two methodologies: these methodologies are in the same class which is the encryption and cryptography class, that's mean these methodologies are using encryption, decryption, public key and hash functions. In section 3.2 a description on how the proposed algorithm will work:

*P2DAP algorithm*

The proposed algorithm is a combination of two methodologies the first one is P2DAP. In which each vehicle will own a number of pseudonyms, the pseudonyms are provided yearly by the DMV. The generated pseudonyms will be hashed using a one-way global key kc, using a hash function called SHA1 [34]. This key will be distributed to all RSUs in the network. Number of bits is selected from these hashed pseudonyms; these selected bits are called "coarse-grained hash value". Afterwards the pseudonyms are grouped into groups according to these bits and called "coarse-grained hash group". Then these new groups are hashed again using one-way key called kf, but this key is not distributed to RSUs. It will be known only by the DMV. Then these groups will be grouped according to the selected bits called "fine-grained group". Each "fine group" is assigned to one vehicle.

Afterwards, the generated pseudonyms will be distributed to vehicles; these pseudonyms are also called "secure plate number" in order to keep the privacy of the drivers no entity can link between the pseudonyms and the owner of the vehicle or the driver of the vehicle. Next the DMV will broadcast the kc to RSUs in the network in order to check whether there is a suspicious event or not.

This algorithm will work on a case that the speed of the vehicles is lower than the speed threshold the speed threshold in this experiment was 40 km/h, which is determined by the street administration according to the street nature and length.

Sybil attack detection is done by the following steps: first the RSU will overhear the communication overhead and hence the exchange messages are signed by the pseudonyms that are generated by the DMV and the kc is known by the RSUs, RSUs will calculate the "coarse grained groups" by using the broadcasted key kc, if two or more pseudonyms are in the same coarse grained group then a suspicious event is occurring, therefore, the RSU will send a report that contains the suspicious pseudonyms. Afterwards, the DMV will check if these pseudonyms are in the same "fine-grained hash group", if yes then a Sybil attack is happening consequently it will send a report to the RSU to terminate the attack. If the suspicious pseudonyms are not in the same "fine grained group", then a false alarm is happening and no action will be done. Figure 2 summarizes the deployment stage of the P2DAP algorithm.

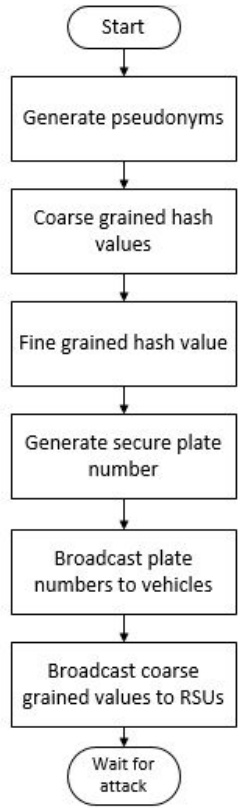

Figure 2: P2DAP algorithm flowchart

*Footprint algorithm*

      The second algorithm is the footprint. This algorithm depends on the trajectory of the vehicle to check whether it is a Sybil node or a genuine node [1].

      To check the trajectory of the vehicle, each vehicle will have a series of link-tags that is obtained by each RSU that is passing by. Thus, the vehicles will have a series of link-tags.

      The Sybil attack detection will be done by checking this series of link-tags, if two vehicles have the same series of link-tags then a Sybil attack is happening, and the RSU will immediately terminate the attack. Otherwise the vehicles are genuine and no Sybil attack is happening.

The deployment stage of this algorithm is done as follows: Each RSU will detect the neighboring RSUs, each RSU will generate a link-tag to the current timestamp, then it will broadcast these link-tags to other RSUs, the broadcasted link-tag is signed by the RSU, afterwards RSUs will send link-tags to TA for authorization, each vehicle will pass by an RSU will obtain a link-tag from the passing by RSU. Figure 3 illustrates this algorithm. This algorithm will always be running therefore the link-tags must be in a serial order to clarify the trajectory of the vehicle.

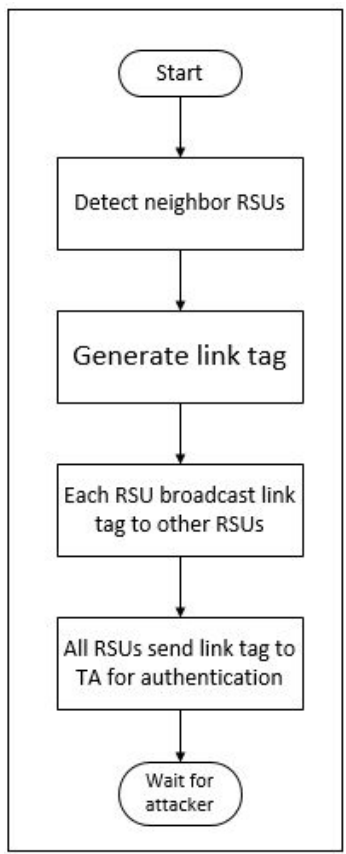

Figure 3: footprint flowchart

### *Hybrid algorithm*

The proposed scheme is a hybrid algorithm between P2DAP and footprint, when the speed is increased over the threshold footprint will be applied, otherwise P2DAP will be applied.

At the beginning, the proposed algorithm must be deployed in the streets. Thus, DMV will yearly generate pseudonyms to all vehicles in the network. Then it will hash these pseudonyms in one-way hash function using a global key kc. Thereafter the hashed pseudonyms will be grouped according to selected bits called "coarse-grained hash values" and these groups are called "coarse-grained groups", afterwards, these groups will be hashed using a local key called kf. Then the global key kc will be distributed to all RSUs in the system. And pseudonyms will be distributed to the vehicles in the network according to the "fine-grained values" each "fine-grained group" will be assigned to one vehicle. Subsequently, each RSU will generate a link-tag that is signed by it, these link-tags are authorized by the TA, and broadcast there link-tags to the neighboring RSUs in order to identify itself and its signature to the other neighboring RSUs in the network.

Figure 4 illustrates the hybrid algorithm. First, it will detect the vehicles on the streets, each vehicle will obtain the deployment information from the nearest RSU, these information will be authorized by the RSUs, afterwards, the RSUs will check the average speed of the vehicles on the street, if the speed is exceeded the threshold which is 40 km/h; the speed threshold could be varied based on several factors such as the maximum speed of the street, the street length and the number of vehicles in the street, the footprint algorithm will run to check if there is a Sybil attack or not, by checking the link-tags for the passing vehicles if they are valid and there are no similarities among them, if there is no similarity among them then there are no attacks detected, otherwise, the RSU will terminate the attack. Thereafter, the proposed algorithm will recheck the streets again to check the average speed of the vehicles. Otherwise the P2DAP algorithm will run to determine whether there is a Sybil attack or not. First the RSU will overhear the message exchanging in the network, each message is signed by the pseudonym of the vehicle. The nearest RSU will hash the existing pseudonyms using the global key kf, if two or more pseudonyms are in the same "coarse-grained hash group", then a suspicious event is occurring, then the RSU will send a report to the DMV to check whether these pseudonyms are in the same "fine-grain group", if yes then a Sybil attack is occurring otherwise, it is a false alarm.

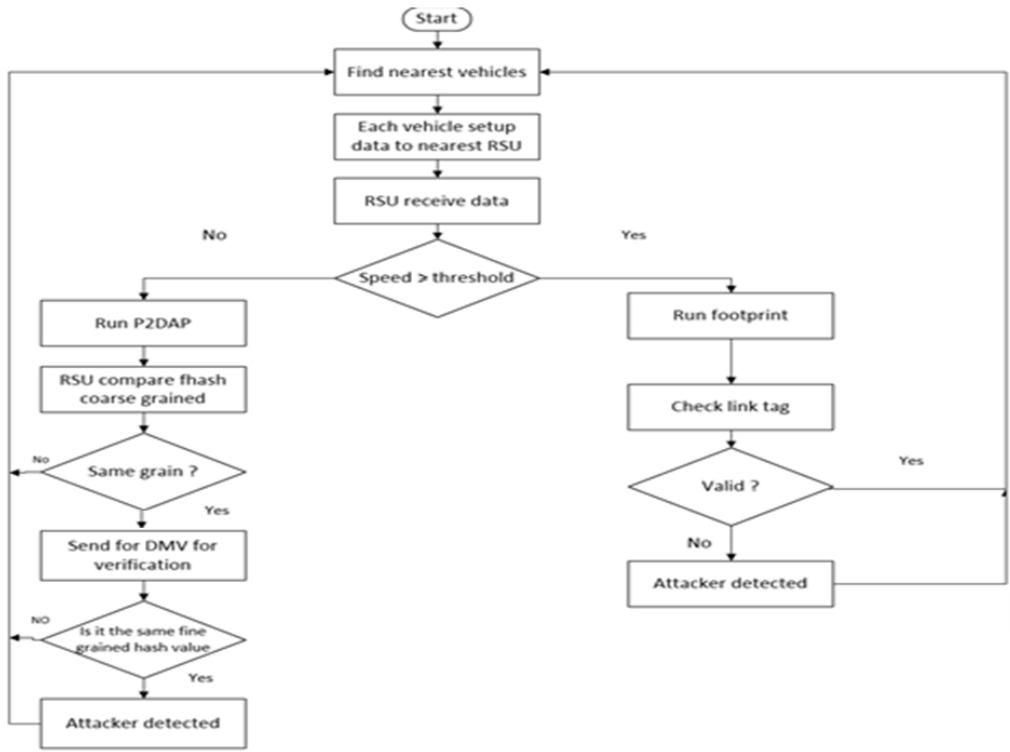

Figure 4: Hybrid algorithm flow chart

*Performance evaluation*

The hybrid algorithm is simulated using ns-2 version 2.35. Ten different scenarios have been generated using MOVE and SUMO tools. The street length is 300m, Simulation time is 900 seconds and the routing protocol is AODV. Details of the simulation parameters are presented in Table I.

Each scenario is simulated using the ns2 simulator and each scenario took 900s. The hybrid algorithm is written using a TCL script. As mentioned before the proposed algorithm is a combination of two algorithms the first one is footprint and the second one is P2DAP. In order to detect if there is a Sybil attack in the network, one of these two algorithms will be running according to the speed condition of the vehicles. Footprint will always be running in order to keep the link-tags in a sequential order. But the proposed algorithm will follow the speed condition to count the number of detected Sybil attackers, if the speed exceeded the threshold which is determined in the performance evaluation to 40 km/h; the reason of choosing this threshold is the length of the street which is 300m, if the streets are longer the speed threshold will increase. The P2DAP algorithm will run otherwise the footprint algorithm will keep running.

Table I Parameters used in the simulation

| Simulation Parameters | Value |
|---|---|
| Simulation time | 900s |
| Routing protocol | AODV |
| Length of road | 300m |
| Number of lanes | 2 |
| Number of RSUs/RSBs | 4 |
| link layer type | LL |
| MAC type | Mac/802_11 |
| Number of attackers | 5 |

The nodes are defined into four types: Authorized vehicles, malicious vehicle (attacker), (RSB in P2DAP, RSU in footprint) and (DMV in P2DAP, TA in footprint). Figure 5 illustrates the simulation map, which is generated using MOVE and SUMO tools.

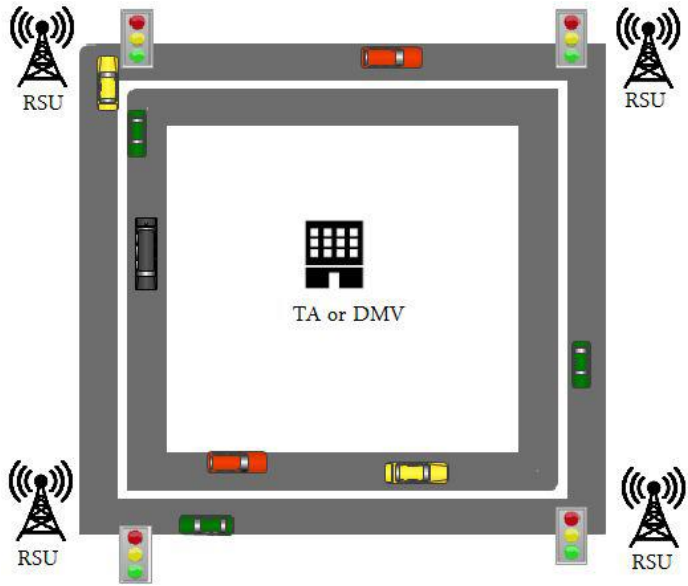

Figure: Simulation map

As it shown in table II, the proposed algorithm is simulated using 10 scenarios, each scenario is simulated on each speed, and then the average of the detection rate result was taken. The detection rate is estimated by collecting the number of detected attackers, each detected attacker will increase the detection rate 20%. The results show that the detection rate of the hybrid algorithm is higher than using the P2DAP and footprint algorithms alone, furthermore the results have a positive relationship with the speed, the P2DAP detection rate is less than the footprint when applying alone. When a hybrid algorithm is applied the detection rate for the Sybil attack is increased, when the speed increases the detection rate is increased. Hence the footprint scheme must be always working; therefore, the link-tag that is obtained to vehicles by RSUs must be in a sequential order, therefore, in the simulation the repeated attackers that are captured by the footprint algorithm while the P2DAP algorithm is running are removed. Figure 6 illustrates the results of simulating the proposed algorithm.

Table II: simulation results

| Speed | Footprint | P2DAP | hybrid |
|---|---|---|---|
| 20km/h | 46% | 42% | 48% |
| 40km/h | 48% | 44% | 50% |
| 60km/h | 50% | 44% | 52% |
| 80km/h | 50% | 48% | 52% |

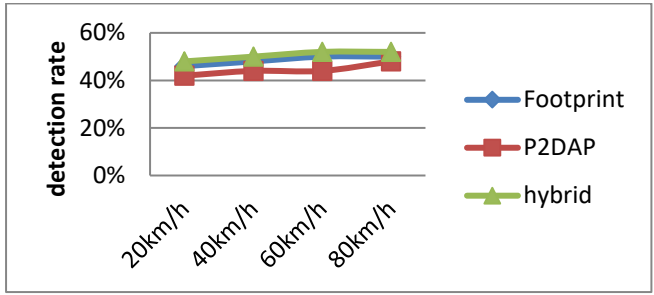

Figure 6: Detection rate for speed varying

**Conclusion**

In this paper a Hybrid algorithm is proposed to detect the Sybil attack. This algorithm is a combination between P2DAP and footprint algorithms, when the speed increases footprint algorithm will be applied to detect the Sybil attack, otherwise the P2DAP will detect the Sybil attack. The results show that the detection rate is increased when applying the hybrid algorithm. The proposed algorithm faces some challenges which are the speed threshold that is not fixed yet and need more research about the streets nature, one other challenge is the algorithm code, which is written by a TCL script, which making the enhancement on the algorithm a bit harder than using a structured language such as C++ or JAVA languages.

**Future work**

In the future, improvements on the proposed algorithm will be done; first of all, the algorithm will be coded using object-oriented language not a TCL script, therefore, the improvement on the algorithm will be applied easily in the future. Also, the optimal speed threshold will be studied based on different factors such as the street nature, the number of nodes in the street and the number of lanes.

**References**

1. Hamdan, S., R.S. Al-Qassas, and S. Tedmori, *Comparative Study on Sybil Attack Detection Schemes.* International Journal of Computers and Technology, 2015. **14**: p. 5869-5876.
2. Dabboussi, A., et al., *Analyzing the reliability for connected vehicles using qualitative approaches and quantitative methods*, in *Safety and Reliability–Safe Societies in a Changing World*. 2018, CRC Press. p. 2603-2610.
3. Qian, Y. and N. Moayeri. *Design of secure and application-oriented VANETs*. in *Vehicular Technology Conference, 2008. VTC Spring 2008. IEEE*. 2008. IEEE.
4. Zhao, Z., et al., *Mobility Prediction-Assisted Over-The-Top Edge Prefetching for Hierarchical VANETs.* IEEE Journal on Selected Areas in Communications, 2018.
5. Newsome, J., et al. *The sybil attack in sensor networks: analysis & defenses*. in *Proceedings of the 3rd international symposium on Information processing in sensor networks*. 2004. ACM.
6. Jain, M. and R. Saxena. *VANET: Security Attacks, Solution and Simulation*. in *Proceedings of the Second International Conference on Computational Intelligence and Informatics*. 2018. Springer.
7. Rawat, A., S. Sharma, and R. Sushil, *VANET: Security attacks and its possible solutions.* Journal of Information and Operations Management, 2012. **3**(1): p. 301.
8. Al-Sultan, S., et al., *A comprehensive survey on vehicular Ad Hoc network.* Journal of network and computer applications, 2014. **37**: p. 380-392.
9. Aboobaker, A.K.K., *Performance analysis of authentication protocols in vehicular ad hoc networks (VANET).* Master of Science Thesis, Department of Mathematics, University of London, September, 2010. **2**.

10. Patel, A. and P. Kaushik, Improving QoS of VANET Using Adaptive CCA Range and Transmission Range both for Intelligent Transportation System. Wireless Personal Communications, 2018. **100**(3): p. 1063-1098.
11. Sabahi, F. The Security of Vehicular Adhoc Networks. in Computational Intelligence, Communication Systems and Networks (CICSyN), 2011 Third International Conference on. 2011. IEEE.
12. Alasmary, W. and W. Zhuang, Mobility impact in IEEE 802.11 p infrastructureless vehicular networks. Ad Hoc Networks, 2012. **10**(2): p. 222-230.
13. Rai, A.K., R.R. Tewari, and S.K. Upadhyay, Different types of attacks on integrated MANETInternet communication. International Journal of Computer Science and Security, 2010. **4**(3): p. 265- 274.
14. Porwal, V. and R. Patel, A survey of VANETs: The Platform for Vehicular Networking Applications. international journal of advanced research in computer engineering & technology, 2014. **3**(8): p.2801-2805.
15. Dhamgaye, A. and N. Chavhan, Survey on security challenges in VANET 1. 2013.
16. Rawat, A., S. Sharma, and R. Sushil, VANET: security attacks and its possible solutions. Journal of
17. Information and Operations Management, 2012. **3**(1): p. 301-304.
18. Hasrouny, H., et al., VANet security challenges and solutions: A survey. Vehicular Communications, 2017. **7**: p. 7-20.
19. Raya, M. and J.-P. Hubaux. The security of vehicular ad hoc networks. in Proceedings of the 3rd ACM workshop on Security of ad hoc and sensor networks. 2005. ACM.19. Grover, J., et al., Performance evaluation and detection of sybil attacks in vehicular ad-hoc networks, in Recent Trends in Network Security and Applications. 2010, Springer. p. 473-482.
20. Haseeb, K., et al., A Survey of VANET's Authentication. Islamia College Peshawar, Pakistan, 2010.
21. Yu, B., C.-Z. Xu, and B. Xiao, Detecting sybil attacks in vanets. Journal of Parallel and Distributed Computing, 2013. **73**(6): p. 746-756.
22. Grover, J., M. Gaur, and V. Laxmi, Sybil Attack in VANETs. Security of Self-Organizing Networks: MANET, WSN, WMN, VANET, 2010: p. 269.

23. Park, S., et al. Defense against sybil attack in vehicular ad hoc network based on roadside unit support. in Military Communications Conference, 2009. MILCOM 2009. IEEE. 2009. IEEE.
24. Guette, G. and B. Ducourthial. On the Sybil attack detection in VANET. in Mobile Adhoc and Sensor Systems, 2007. MASS 2007. IEEE International Conference on. 2007. IEEE.
25. Yan, G., S. Olariu, and M.C. Weigle, Providing VANET security through active position detection. Computer Communications, 2008. **31**(12): p. 2883-2897.
26. Viswanath, B., et al. Exploring the design space of social network-based sybil defenses. In Communication Systems and Networks (COMSNETS), 2012 Fourth International Conference on. 2012. IEEE.
27. Xiao, B., B. Yu, and C. Gao. Detection and localization of sybil nodes in VANETs. in Proceedings of the 2006 workshop on Dependability issues in wireless ad hoc networks and sensor networks. 2006. ACM.
28. Yan, G., et al. Active position security through dynamically tunable radar. in Mobile Adhoc and Sensor Systems (MASS), 2010 IEEE 7th International Conference on. 2010. IEEE.
29. Rahbari, M. and M.A.J. Jamali, Efficient detection of sybil attack based on cryptography in VANET. arXiv preprint arXiv:1112.2257, 2011.
30. Studer, A., et al. TACKing together efficient authentication, revocation, and privacy in VANETs. In Sensor, Mesh and Ad Hoc Communications and Networks, 2009. SECON'09. 6th Annual IEEE Communications Society Conference on. 2009. IEEE.
31. Chang, S., et al., Footprint: Detecting sybil attacks in urban vehicular networks. Parallel and Distributed Systems, IEEE Transactions on, 2012. **23**(6): p. 1103-1114.
32. Park, S., et al., Defense against Sybil attack in the initial deployment stage of vehicular ad hoc network based on roadside unit support. Security and Communication Networks, 2013. **6**(4): p. 523- 538.
33. Zhou, T., et al., P2DAP—Sybil attacks detection in vehicular ad hoc networks. Selected Areas in Communications, IEEE Journal on, 2011. **29**(3): p. 582-594.

34. Rogaway, P. and T. Shrimpton. Cryptographic hash-function basics: Definitions, implications, and separations for preimage resistance, second-preimage resistance, and collision resistance. in Fast Software Encryption. 2004. Springer.